\documentclass[sigconf]{acmart}

\usepackage{listings}
\usepackage{nicefrac}

\copyrightyear{2025}
\acmYear{2025}
\setcopyright{rightsretained}

\acmISBN{978-1-4503-XXXX-X/18/06}

\begin{document}

\title{Debugging Without Error Messages}
\subtitle{How LLM Prompting Strategy Affects Programming Error Explanation Effectiveness}

\author{Audrey Salmon}
\orcid{0009-0000-5511-9904}
\affiliation{
  \institution{Department of Computer Science}
  \institution{University of North Carolina at Chapel Hill}
  \city{Chapel Hill}
  \country{USA}
}
\email{audrey.b.salmon@gmail.com}

\author{Katie Hammer}
\orcid{0009-0002-0668-5349}
\affiliation{%
  \institution{School of Computer Science}
  \institution{North Carolina State University}
  \city{Raleigh}
  \country{USA}
}
\email{kahammer@ncsu.edu}

\author{Eddie Antonio Santos}
\orcid{0000-0001-5337-715X}
\affiliation{%
  \institution{School of Computer Science}
  \institution{University College Dublin}
  \city{Dublin}
  \country{Ireland}
}
\email{eddie.santos@ucdconnect.ie}

\author{Brett A. Becker}
\orcid{0000-0003-1446-647X}
\affiliation{
  \institution{School of Computer Science}
  \institution{University College Dublin}
  \city{Dublin}
  \country{Ireland}
}
\email{brett.becker@ucd.ie}

\newcommand{\citationneeded}{\textsuperscript{\color{blue} [citation needed]}}

\begin{abstract}
Making errors is part of the programming process---even for the most seasoned professionals.
Novices in particular are bound to make many errors while learning.
It is well known that traditional (compiler/interpreter) programming error messages have been less than helpful for many novices and can have effects such as being frustrating, containing confusing jargon, and being downright misleading.
Recent work has found that large language models (LLMs) can generate excellent error explanations, but that the effectiveness of these error messages heavily depends on whether the LLM has been provided with context---typically the original source code where the problem occurred. 
Knowing that programming error messages can be misleading and/or contain jargon that serves little-to-no use (particularly for novices) we explore the reverse: what happens when GPT-3.5 is prompted for error explanations on just the erroneous source code itself---original compiler/interpreter produced error message excluded.
We utilized various strategies to make more effective error explanations, including one-shot prompting and fine-tuning.
We report the baseline results of how effective the error explanations are at providing feedback,
as well as how various prompting strategies might improve the explanations' effectiveness.
Our results can help educators by understanding how LLMs respond to such prompts that novices are bound to make,
and hopefully lead to more effective use of Generative AI in the classroom.
\end{abstract}



\maketitle

\section{Introduction}%
\label{sec:introduction}

Students and professionals alike frequently make mistakes when writing code.
Some of these mistakes completely grind the programming process to a halt,
resulting in diagnostics called programming error messages (PEMs).
Much ink has been spilled lamenting the quality of PEMs throughout the decades~\cite{becker2019compiler}---%
often labeled as ``terse'', ``cryptic'', and ``misleading''---%
but recently, a new alternative to traditional error messages has been made available.

Generative AI tools such as ChatGPT have suddenly shaken up multiple fields including software development and computing education.
Prior work has shown that large language models (LLMs), like the one used in ChatGPT,
can provide acceptable explanations for programming error messages~\cite{leinonen2023using,santos2023always,widjojo2023addressing}.
A common theme in these works is that adding source code context drastically improves the quality of generated explanations.

The present work asks a curious question:
\emph{How effective is the feedback generated by LLMs when the original programming error message is omitted entirely?}
This work not only establishes a baseline for feedback without error messages,
but attempts to improve on this baseline using a variety of \emph{prompting strategies}.
We compare one-shot prompting---providing an example of the desired feedback in the prompt---%
with fine-tuning---%
a more involved process that requires modifying the language model's weights using additional training examples. 

This work hopes to offer insights to educators,
both with regards to the role of feedback when resolving programming errors,
and to the use of generative AI in the classroom.

\subsection{Research questions}

We are guided by the following research questions:
\begin{enumerate}
    \item[RQ1] How effective are LLM-generated error message explanations that omit the original error message from the prompt? 
    \item[RQ2] How does prompting strategy affect various aspects of LLM-generated error explanations, including\ldots
    \begin{itemize}
        \item \ldots{}the \emph{accuracy} of error explanations?
        \item \ldots{}the \emph{relevancy} of error explanations?
        \item \ldots{}the \emph{verbosity} of error explanations?
    \end{itemize}
    \item[RQ3] What trade-offs are there between different prompting strategies in the context of generating error explanations?
\end{enumerate}

\subsection{Contributions}

With this work, we provide empirical evidence that:

\begin{itemize}
    \item Prompting GPT 3.5 for feedback without the original error message produces roughly 2--3 useful responses for every misleading explanation generated.
    \item Alternative prompting strategies (one-shot and fine-tuning) do not appreciably increase the accuracy of LLM generated error explanations.
    \item One-shot prompting and fine-tuning produce fewer instances of distracting, extraneous information in the generated error explanations.
    \item Fine-tuning generates error explanations that are more concise and on-topic than the other prompting strategies.
\end{itemize}

\section{Background and Related Work}%
\label{sec:relatedwork}

Although the present work proposes to remove programming error messages from the process of debugging,
it is useful to understand why such an idea would seem reasonable in the first place.

\subsection{Programming Error Messages}
Programming error messages (PEMs) are the (usually textual) diagnostic messages that are generated
when an unrecoverable error is detected while programming---either at compile-time, or while the program is running.
They are one of the primary forms of feedback that both novice and professionals programmers alike receive while coding.
PEMs have had a long and unfortunate history of being perceived as unhelpful~\cite{becker2019compiler}.
They often contain technical jargon unfamiliar to novices and have a penchant for making misleading suggestions~\cite{dy2010detector,kohn2019error,marceau2011mind}.
PEMs have such a bad reputation that the present work proposes removing them altogether---%
however, prior work has found that any error message is better than no error message at all~\cite{shneiderman1982system}.

Recent work has tried to understand the factors that make PEMs understandable. 
These factors suggest that appropriate feedback should be short and succinct,
use simple vocabulary without any jargon, and be written in clear sentences~\cite{denny2021designing}.
Unfortunately, these guidelines are difficult to action within compilers because producing
succinct and precise feedback requires a level of understanding of the programmer's intent.
Programming \emph{errors} and programming error \emph{messages} are not the same thing~\cite{mccall2014meaningful};
producing useful feedback---especially for novices---requires more careful consideration of what mistakes are likely and which are difficult to overcome while programming~\cite{mccall2019new}. 
That said, statistical modeling has shown that programmers' intent is largely predictable~\cite{hindle2012naturalness},
enabling efforts to further use statistical modeling to generate feedback as an alternative to conventional programming error messages~\cite{campbell2014syntax,santos2018syntax}.

\subsection{Generative AI and Large Language Models}
The world has been rocked by the flood of easily accessible Generative AI applications.
Approachable chatbot interfaces like ChatGPT allow laypeople to interact with the world's largest AI models
using the same social scripts that they would use in human-human interaction~\cite{nass2000machines}.
Large language models (LLMs) such as GPT-3.5 (the LLM introduced with ChatGPT in November 2022),
have had a profound impact on computing education in a relatively short time.
Early work found that LLMs can solve most CS1~\cite{finnieansley2022robots} and CS2 problems~\cite{finnieansley2023myai},
raising concerns that automated assessment could soon be trivially circumvented.
Indeed, the computing education community is struggling to integrate generative AI in teaching~\cite{lau2023ban}.

\subsection{Using LLMs to generate programming error feedback}
Prior work has used LLMs to tackle the problem of frustrating programming error messages.
One work using an early, code-oriented LLM, found that it could produce correct explanations for error messages in 48\% of cases,
but only 33\% of all generated explanations had correct fixes.
Later work, using more capable models like GPT-4, found that LLMs could generate
responses with up to 99\% correct explanations and 83\% correct fixes~\cite{santos2023always},
and up to 100\% ``useful'' responses~\cite{widjojo2023addressing}.
Most notably, all of the previous studies showed that programming feedback was most effective when it included the original erroneous source code as part of the prompt.
These studies have also shown that only providing the programming error message to LLMs often produces vague, ``technically correct'' responses,
instead of targeted, actionable feedback.
Other work has used LLMs in clever ways to generate feedback for syntax errors,
without the use of the original programming error message~\cite{phung2023generating}.

\section{Methodology}%
\label{sec:methodology}

To compare the three different prompting strategies (baseline, one-shot, fine-tuned),
we first collected 100 erroneous student programs from the TigerJython dataset.
We manually wrote error explanations for 60 programs (\autoref{sec:manualexplanations}),
holding out the remaining 40 for evaluation.
We prompted GPT-3.5 (\autoref{sec:promptingstrategies}),
incorporating the manual explanations into our one-shot and fine-tuned prompting strategies.
Finally, we manually evaluated the three prompting strategies using the remaining 40 programs,
rating based on criteria derived from prior work (\autoref{sec:evaluation}).

\subsection{Collecting erroneous student programs}%
\label{sec:dataset}

We sampled 100 erroneous student programs from the TigerJython 2022 ProgSnap2 database~\cite{kohn2020tell}.
TigerJython is a introductory programming environment, aimed at secondary school students.
The programming language is a modified version of Jython, itself a dialect of Python 2.
Though it resembles Python 2, TigerJython features a few extended syntactic constructs that are not present in either Python 2 or Python 3.
The fact that TigerJython code resembles but is not identical to Python 3
became an issue when obtaining responses from GPT-3.5 (see \autoref{sec:implications}).

To find suitable erroneous code, we used the following criteria:

\begin{description}
    \item[error $\rightarrow$ fixed]  The erroneous program must originate from a pair of events where the code raised an error in one event, and then ran successfully in the subsequent event.
    \item[$\leq$ 20 lines] The erroneous program consists of at most 20 lines of code.
    \item[Python 2] The fixed program uses valid Python 2.7 syntax.
    \item[1 error] The erroneous program has exactly one programming error that results in an error message being emitted, either at compile-time or at runtime.
\end{description}

The first three criteria were used to automatically filter the dataset.
The last criterion (exactly one programming error) required manual analysis.
As such, the first author randomly sampled from the filtered dataset until 100 programs were found to have exactly one programming error.
Of the 100 erroneous programs selected,
60 programs were used to write manual error explanations (\autoref{sec:manualexplanations}),
while the remaining 40 were held out for the evaluation (\autoref{sec:evaluation}).

\paragraph{Censored strings}
For privacy reasons, programs in the dataset had ``censored'' string literals such that some text would be replaced with a series of X's.
For example, \lstinline{setColor("Yellow")} would show in the dataset as \lstinline{setColor("Xxxxxx")}.
In preliminary testing, we found that GPT-3.5 would suggest changing the string to a valid color name,
which was not the intended programming error to correct.
However, the strings were only censored in the source code---the error messages stored in the dataset (inadvertently) revealed uncensored string literals.
In these cases, we manually changed the censored string literal back to a ``reasonable'' string literal
(e.g., \lstinline{"Xxxxxx"} $\rightarrow$ \lstinline{"Yellow"}) to prevent the LLM from suggesting this as the error.

\subsection{Creating manual error explanations}%
\label{sec:manualexplanations}

We manually created error explanations for 60 erroneous programs.
Two authors split the task, writing 30 explanations each.
To better understand the students' intent, authors examined both the erroneous program and the student's fix to the problem.
Using this information, we wrote concise error explanations in the following consistent format, featuring:
\begin{enumerate}
    \item One or two complete sentences explaining the problem.
    \item One or two complete sentences explaining the fix.
    \item A minimal example of correct source code.
\end{enumerate}

\noindent
For example:
\begin{quote}
    Running the provided code results in an error because the \lstinline{forward()} function needs to include a numerical value.
    To fix the problem, give \lstinline{forward()} a value.
    For example, \lstinline{forward(30)}.
\end{quote}

In cases where the fix contained the example source code in its entirety, we omitted the (now redundant) example to make each message more succinct.
For example:
\begin{quote}
    Running the provided code results in an error because the \lstinline{maketurtle()} function needs to have a capital T.
    To fix the problem, change it to \lstinline{makeTurtle()}.
\end{quote}

\subsection{Models and prompting strategies}%
\label{sec:promptingstrategies}

We compared three different prompting strategies:
the baseline,
prompting a model with one example of the desired error explanation (one-shot),
and prompting a model fine-tuned on manual error explanations.
In all cases, we used OpenAI's gpt-3.5-turbo-1106 model.
As of this writing, GPT-3.5 is the latest model family from OpenAI that is generally available for fine-tuning.\footnote{
    \url{https://platform.openai.com/docs/models/gpt-3-5-turbo}
}
We prompted using OpenAI Playground,\footnote{
    \url{https://platform.openai.com/playground/chat?models=gpt-3.5-turbo-1106}
} which allowed us to directly control hyperparameters.
In particular, we used a temperature of 0.0 when obtaining responses,
as this was found to be the most effective in prior work~\cite{leinonen2023using}.
For all three prompting strategies, we provided the full, unabridged erroneous source code as the user message.

\subsubsection{Baseline}%
\label{sec:baseline}

The baseline (control) was to prompt gpt-3.5-turbo-1106 using the following system message:
\begin{quote}
    Provide a plain English explanation of why running the Python 2 code causes an error and how to fix the problem. Do not output the entire fixed source code.
\end{quote}
\noindent
This message is based on prompt \#1 from \citet{leinonen2023using}, with a few notable changes:
we explicitly specify that the code is in Python 2 because we found in preliminary testing that the models would suggest to fix syntactic elements to make them compatible with Python 3 rather than explain the true programming error.
In addition,
OpenAI's language models seem to consistently reproduce the \emph{entire} source code with the problem fixed when providing suggestions~\cite{santos2023always}.
For our purposes, this is completely unnecessary, so we included wording in the system message to omit outputting the entire fixed program.

\subsubsection{One-shot prompting}
A way to improve the output from an LLM is to provide one or more examples of the desired output in the prompt.
This is called \emph{few-shot prompting}~\cite{openai2024prompt}.\footnote{%
    The baseline (not using examples in the prompt) is also known as \emph{zero-shot learning}.%
}
This is a relatively low-effort method to potentially increase the effectiveness of an LLM's output.
In our case, we evaluated the case where exactly one example is added to the prompt (i.e., one-shot prompting).
For each erroneous program,
we augmented the baseline's system message by concatenating it with the phrase ``For
example:'' followed by an error explanation randomly sampled from our set of 60 manually
written explanations (\autoref{sec:manualexplanations}).

\subsubsection{Fine-tuned model}
A more involved method of improving model output is taking a pretrained model (the `P' in `GPT') 
and adjusting some of the model's weights with further training examples.
This process, known as \emph{fine-tuning}, turns a general-purpose model into a specialized model. A fine-tuned model has the same architecture as the base model, but has more targeted or specialized capabilities.
Research has indicated that large language models require relatively few samples for fine-tuning, requiring as few as 100 examples to match the performance of a model trained from scratch on $100\times$ more data~\cite{howard2018universallanguagemodelfinetuning}.
OpenAI recommends 50 to 100 training examples for fine-tuning GPT-3.5, but notes that ``the right number varies greatly based on the exact use case''~\cite{openai2024finetuning}.

Using OpenAI Playground, we took the baseline model (\autoref{sec:baseline}),
and fine-tuned it with all 60 handwritten explanations (\autoref{sec:manualexplanations}) with their respective erroneous programs.
For each training example, we provided our manually written explanation as the assistant message,
leaving both the system message and user message as they would be in the baseline.
We let epochs, batch size, and LR multiplier all be automatically set,
which resulted in the system using 3 epochs, a batch size of 1, and an LR multiplier of 2.
In all, fine-tuning took just under 8 minutes.
During the evaluation, the fine-tuned model was prompted in exactly the same way as the baseline model.

\subsection{Evaluation}%
\label{sec:evaluation}

Three authors rated the LLMs' responses on two axes:
\textbf{feedback quality} and \textbf{extraneous information}.

Labels for feedback quality were adapted from \citet{mahajan2020recommending} via \citet{widjojo2023addressing}.
We partially quote Table 3 from the latter work here:
\begin{quote}
    \begin{description}
        \item[Instrumental (I)] Response perfectly targets underlying cause of error and provides a clear action on how to fix.
        \item[Helpful (H)] Response provides general but not exact help. [\ldots]
        \item[Misleading (M)] Response does not provide a clear direction on how to fix the issue and/or causes confusion. [\ldots]
    \end{description}
\end{quote}

A critique of \citet{widjojo2023addressing} is that in the original formulation, ``misleading'' seems to cover two axes:
that the \emph{primary} feedback is misleading, or that the response contains extra information that may further mislead.
Indeed, preliminary work revealed that GPT would occasionally produce correct responses with extraneous information that did not directly address the error.
This extraneous information would range from being technically correct (``Additionally, you may need to make some adjustments to the function calls and syntax to ensure compatibility with Python 3.'') or incorrect and misleading
(``To fix the problem, you can use Python 3 instead, as the \lstinline{gturtle} library is designed for Python 3''
while commenting on correct usage of \lstinline{gturtle} in TigerJython).
As a result, we assessed the message on an additional axis: whether the response contained \textbf{extraneous information}.
For this, the first three authors collectively agreed on codes for labeling extraneous information:

\begin{description}
    \item[No extraneous information] The entirety of the response is relevant to diagnosing and fixing the one programming error.
    \item[Correct, but extraneous information] The response contains irrelevant information, but it is factually accurate or useful.
    \item[Incorrect extraneous information] The response contains irrelevant information that may further mislead.
\end{description}

\paragraph{Evaluation metrics}
As in \citet{widjojo2023addressing}, we report summary metrics (\autoref{tab:ihm-definitions})
that are intended to give an idea of the overall feedback quality of each prompting strategy.

\begin{table}[hb]
    \centering
    \caption{Metrics for evaluating feedback quality, adapted from \citet{widjojo2023addressing}.}%
    \label{tab:ihm-definitions}
    \begin{tabular}{@{}llp{4.5cm}@{}}
        \toprule
         Metric &  Definition & Description \\
        \midrule
         I-Score &  $\frac{I}{I + H + M} \times 100\%$ & What proportion of all responses are instrumental (highest quality)? \\
         IH-Score &  $\frac{I + H}{I + H + M} \times 100\%$ & What proportion of all responses are useful (instrumental or helpful)? \\
         M-Score &  $\frac{M}{I + H + M} \times 100\%$ & What proportion of all responses are misleading? \\
         IH:M ratio & $(I + H):M$ & How many instrumental/helpful responses are there for every misleading response produced? \\
        \bottomrule
    \end{tabular}
\end{table}

\paragraph{Protocol}
Before rating commenced, the raters convened in person and collectively agreed on the interpretation of all labels.
Once all 120 LLM responses were generated ($40$ erroneous programs $\times$ $3$ prompting strategies),
the raters provided an initial rating of every response, with respect to the erroneous program and the student's intention.
After this initial rating, the authors reconvened and discussed special and marginal cases.
There was only one case in which all three authors completely disagreed.
After brief discussion, the ratings were revised such that there was a majority opinion for each response.

\begin{table*}[tbh]
  \centering
  \caption{Summary of majority ratings (40 erroneous programs per row).}%
  \label{tab:results}
  \begin{tabular}{@{}lrrrrrrrrrr@{}}
     \toprule
               & \multicolumn{7}{c}{Feedback quality}                    & \multicolumn{3}{c}{Extraneous information} \\
                 \cmidrule(lr){2-8}                                        \cmidrule(lr){9-11}
               & I & H & M & I-score  & IH-Score  & M-Score  & IH:M      & None & Correct & Incorrect \\
     \midrule      
     Baseline   & 17 & 11 & 12 & 45.0\% & 70.8\% & 29.2\% & 2.43:1 & 33 & 6 & 1 \\
     One-shot   & 21 & 9  & 10 & 50.8\% & 76.7\% & 23.3\% & 3.29:1 & 37 & 2 & 1 \\
     Fine-tuned & 21 & 8  & 11 & 52.5\% & 71.7\% & 28.3\% & 2.53:1 & 40 & 0 & 0 \\
      \bottomrule
\end{tabular}
\end{table*}

\section{Results}%
\label{sec:results}

\autoref{tab:results} summarizes the results of from the three raters.
We report the raters' majority opinion for \textbf{I}ns\-tru\-men\-tal/\textbf{H}elp\-ful/\textbf{M}is\-lead\-ing,
and the presence of extraneous information,
such that these totals add up to 40 (the number of erroneous programs);
however, the other measures are calculated on the raw frequencies from all raters.
Fliess' $\kappa$ for the three raters was calculated on the two axes separately.
We obtained $\kappa = 0.725$ for feedback quality and $\kappa = 0.809$ for extraneous information,
which may be interpreted as significant and almost-perfect agreement~\cite{landis1977measurement}, respectively.

\paragraph{Feedback quality} The feedback quality obtained on the baseline is remarkable:
without providing the original programming error message to GPT-3.5, the baseline results 
in 70.8\% of its responses as being rated useful and 45.0\% \emph{perfectly} explaining and fixing the error.
Compare this to the 11\% perfect fix rate reported in prior work~\cite{santos2023always} that prompted using only the programming error message using a more capable model (GPT-4).

We conducted a $\chi^2$ test to determine whether feedback quality is affected by prompting strategy,
and found no statically significant difference ($p=0.53$).
Thus, we conclude that prompting strategy---%
whether using an ordinary prompt, a prompt with an example of desired output, or even a model fine-tuned with desired feedback---%
does not affect the proportion of instrumental, helpful, or misleading explanations produced by GPT-3.5.
One can expect between 2.43--3.29 useful explanations for every misleading explanation obtained.

\paragraph{Extraneous information}
A $\chi^2$ test revealed that there is a statistically significant ($p<0.001$) effect between prompting strategy and the presence of extraneous information in the response.
Notably, while the baseline and one-shot strategies produced explanations that were judged as sometimes containing extraneous information,
all three raters unanimously agreed that every response from the fine-tuned model was completely devoid of extraneous information.
That is, the fine-tuned model's explanations were 100\% on topic.

\begin{figure}[tbh]
    \centering
    \includegraphics[width=\columnwidth]{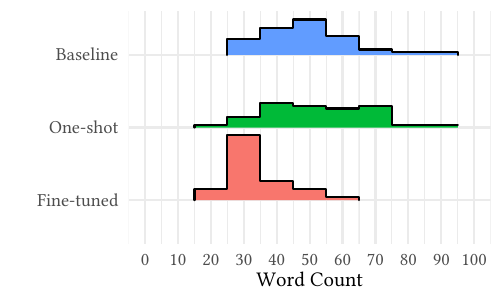}
    \caption{Histograms of word count by prompting strategy.}%
    \label{fig:wordcount}
    \Description{%
        Three stacked histograms, indicating the number of words per explanation, grouped by prompting strategy: %
        from top-to-bottom, the baseline results, the one-shot results, then the fine-tuned results. %
        The baseline histogram is roughly normally distributed, but right-tailed, starting from a minimum of 25 words, %
        to a maximum of 95 words, with a mode of 45--55 words. %
        The one-shot distribution is mildly bimodal, with a minimum of 15 words, a maximum of 95 words, with a mode at 35--45 words, %
        and a less prominent mode at 65--75 words. %
        The fine-tuned distribution is last, with a notably smaller range than the previous two distributions: %
        a minimum of 15 words and a maximum of 75 words, with a prominent mode at 25--35 words per explanation.%
    }
\end{figure}

\paragraph{Message length}
We counted the number of words in each message, where a ``word'' is defined by calling Python's \lstinline{str.split()} method on each message and counting length of the resultant list.
\autoref{fig:wordcount} shows the distributions of word counts, grouped by prompting strategy.
A Kruskal-Wallis rank sum test reveals that the differences between distributions is statistically significant ($p<0.001$).
Post hoc paired Wilcoxon rank sum tests show that statistical significant differences exist between the distribution
of word counts of the fine-tuned messages against the baseline ($p<0.001$) and the one-shot prompting strategy ($p<0.001$);
further one-tailed tests reveal that the fine-tuned messages are shorter than both the baseline ($p<0.001$) and the one-shot strategy ($p<0.001$).
We conclude that explanations obtained from the fine-tuned model are less loquacious than explanations obtained from the alternatives.
However, we found that there is no statistically significant difference between the distributions of word counts between the baseline and one-shot strategies ($p = 0.11$).

\section{Discussion}

\subsection{RQ1: How effective are LLM-generated error message explanations that omit the original error message from the prompt?}
Our results show that one can expect roughly 2--3 useful error explanations for every misleading error explanation generated by GPT 3.5
when the original programming error message is completely absent from the prompt.
This is not too surprising,
as prior work has shown that the presence of source code context dramatically improves an LLM's ``perfect fix'' rate of error explanations from 11\% to 79\%~\cite{santos2023always}.
Indeed, early work on applying various automated approaches to fixing students' syntax errors did not use the error message at all~\cite{santos2018syntax,gupta2017deepfix,campbell2014syntax}.
When a later work finally incorporated Java's compiler error messages into their AI tool, they saw only a modest 2.7\% improvement in fix rates~\cite{ahmed2022synshine}.
They also claimed that their model learned to \emph{ignore} the fix hinted at in the message.
This puts in to question the utility of traditional programming error messages as a debugging tool---at least for novice programmers.

\subsection{RQ2: How does prompting strategy affect various aspects of LLM-generated error explanations?}
Prompting strategy had an effect on the presence of extraneous information, and in the case of the fine-tuned messages, the length of the response.
However, prompting strategy did not seem to affect the veracity of the error explanations produced.
It is possible that using a larger model, trained on more data would be more effective as has been tried in other studies~\cite{mozannar2024realhumaneval,santos2023always}.
However, we will note that prior work has found that, when it comes to using LLMs to solve programming tasks, ``further gains in benchmark performance do not necessarily translate into equivalent gains in human productivity''~\cite{mozannar2024realhumaneval}.
The aspects that make an LLM's response actionable may be more complex than whether the response simply contains the correct answer within it.

GPT-3.5's responses did not contain much extraneous information overall, but prompting strategy did seem to lower it---especially using the fine-tuned model.
A lack of extraneous information does not necessarily imply a shorter response, as the one-shot strategy's responses were not any shorter than the baseline's messages overall.
Providing just one example of the desired LLM response may not be enough to produce shorter, more relevant responses;
however, if producing shorter, more relevant responses is a priority, then fine-tuning is more effective.

\subsection{RQ3: What trade-offs are there between different prompting strategies in the context of generating error explanations?}

There is a varying amount of effort required (both for students and instructors) for using all three prompting strategies,
increasing from the baseline strategy, to one-shot/few-shot prompting, to the most time consuming: fine-tuning.
Since the veracity of the feedback does not see any appreciable improvements when using prompting strategies that require more effort,
it may be wiser to use a more capable model, if available.

However, the usability of an error explanation may not just be a function of whether it gives a correct answer,
but also whether it can produce a \emph{concise} message.
For this, it seems that fine-tuning is the most effective strategy overall.
From our experience, fine-tuning is not an overly onerous task:
it took two authors less than one working day to create 30 error explanations each,\footnote{
    An embarrassingly parallel task.
} and only 8 minutes to fine-tune the model proper.
We consider this to be a relatively small time investment for worthwhile results.

The real challenge in fine-tuning is finding \emph{diverse} examples of students' programming errors.
To properly train models, it is imperative that training data have plentiful and varied examples,
and not feature a \emph{class imbalance} where some categories of programming errors occur disproportionately.
This, however, is always the case since students' programming errors are found in a long-tail distribution~\cite{mccall2014meaningful}.
That is, the most common programming errors are extremely frequent, and the least common programming errors are vanishingly rare.
In our sample of 60 programming errors, we found 15 cases (25\%) where the error was a misspelling of a variable or function name,
and 6 cases (10\%) where the mistake was forgetting to initialize the environment.
These imbalances may have caused our fine-tuned model to overfit and thus, not improve in its feedback quality rating.
It is worthwhile to note that programming \emph{errors} are not the same as programming error \emph{messages}:
comparing the distributions of both reflects this~\cite{mccall2014meaningful}.
Thus,
care must be taken when creating datasets of programming errors for the purpose of fine-tuning models
such that there are examples of broad and diverse programming \emph{errors}---not programming error messages.

\subsection{Implications for pedagogy}%
\label{sec:implications}

Minimizing extraneous information in error explanations may be invaluable due to a reduction in extraneous cognitive load for the student.
This is especially true in the context of novice programmers using
pedagogically-oriented language dialects, as is the case with TigerJython.
Since general-purpose large language models like GPT-3.5 have been trained on code coming from a wide variety of domains,
its extra hints may not be helpful or even distracting when using it for highly-specialized teaching languages like TigerJython.
We noticed that GPT-3.5 had a fixation on making the code compatible with Python 3.
Of the 10 cases of extraneous information that we labeled,
8 were messages either stating that the code was not compatible with Python 3,
or suggesting how to change the code to be compatible with Python 3.
None of these suggestions would be relevant to students who were debugging TigerJython (Python 2) code.
It seems that LLMs put students who use pedagogically-oriented languages at a disadvantage because the LLMs' output skews heavily towards the norms of mainstream programming languages and professional software development practices.

An opportunity that this work presents is an exercise in which students reflect on what kind of feedback they require when debugging.
Before debugging with an LLM such as ChatGPT,
students could be asked to craft examples of the structured feedback that they would like to receive,
much like the template defined in \autoref{sec:manualexplanations}.\@
Students would then use their example explanations as part of a one-shot or few-shot prompt to fix a novel programming error,
and reflect on the effectiveness of the resultant error explanation.

\subsection{Limitations}

As with other studies that evaluate LLMs on synthetic benchmarks,
the true test of how helpful LLMs are is demonstrated in how students \emph{actually} use the output in practice.
Related work has shown that promising results in benchmarks do not necessarily translate to promising results in practice~\cite{mozannar2024realhumaneval,prather2024widening,nguyen2024beginning,vaithilingam2022expectation,santos2024not}.

Another limitation is our dataset: we trimmed down the dataset for practical purposes:
namely, evaluation was less onerous for the raters if the programs were short (20 lines or fewer),
and if it was relatively simple to assess whether the \emph{one} programming error was fixed.
In reality, students' programs routinely contain several programming errors simultaneously, and students are not limited to 20 line programs.
Additionally, our ``uncensoring'' of string literals means that some of the erroneous programs we used were not, strictly speaking, identical to the ones that students actually wrote.

\section{Conclusion}
\balance{}

We have demonstrated that, when prompting GPT-3.5 \emph{without} programming error messages,
one can conservatively expect 2--3 useful error explanations for every misleading response. 
Additionally, prompting strategy does not appreciably change the accuracy of the generated error explanations,
but it may at least make the explanations shorter and more focused.
This work adds to the growing pile of evidence that suitable programming error feedback is more reliant on the erroneous source code context than the resultant error message.
More broadly, we hope instructors focus on the causes and resolutions to underlying \emph{programming errors} rather than \emph{programming error messages},
regardless of their use of generative AI.
If GenAI tools such as ChatGPT are introduced in the classroom,
we suggest that it is a better use of time to focus on explaining the underlying programming errors rather than prompting chatbots to explain programming error messages in isolation.
We hope this work better equips educators on how to effectively utilize LLMs in the classroom and helps establish realistic expectations regarding the capabilities of the now ubiquitous generative AI tools.

\begin{acks}
 We are indebted to Tobias Kohn for providing the TigerJython dataset and for his kind advice.
\end{acks}

\bibliographystyle{ACM-Reference-Format}
\bibliography{references}


\begin{thebibliography}{32}


\ifx \showCODEN    \undefined \def \showCODEN     #1{\unskip}     \fi
\ifx \showDOI      \undefined \def \showDOI       #1{#1}\fi
\ifx \showISBNx    \undefined \def \showISBNx     #1{\unskip}     \fi
\ifx \showISBNxiii \undefined \def \showISBNxiii  #1{\unskip}     \fi
\ifx \showISSN     \undefined \def \showISSN      #1{\unskip}     \fi
\ifx \showLCCN     \undefined \def \showLCCN      #1{\unskip}     \fi
\ifx \shownote     \undefined \def \shownote      #1{#1}          \fi
\ifx \showarticletitle \undefined \def \showarticletitle #1{#1}   \fi
\ifx \showURL      \undefined \def \showURL       {\relax}        \fi
\providecommand\bibfield[2]{#2}
\providecommand\bibinfo[2]{#2}
\providecommand\natexlab[1]{#1}
\providecommand\showeprint[2][]{arXiv:#2}

\bibitem[Ahmed et~al\mbox{.}(2022)]%
        {ahmed2022synshine}
\bibfield{author}{\bibinfo{person}{Toufique Ahmed}, \bibinfo{person}{Noah~Rose
  Ledesma}, {and} \bibinfo{person}{Premkumar Devanbu}.}
  \bibinfo{year}{2022}\natexlab{}.
\newblock \showarticletitle{{{SynShine}}: {{Improved Fixing}} of {{Syntax
  Errors}}}.
\newblock \bibinfo{journal}{\emph{IEEE Transactions on Software Engineering}}
  \bibinfo{volume}{49}, \bibinfo{number}{4} (\bibinfo{year}{2022}),
  \bibinfo{pages}{2169--2181}.
\newblock


\bibitem[Becker et~al\mbox{.}(2019)]%
        {becker2019compiler}
\bibfield{author}{\bibinfo{person}{Brett~A. Becker}, \bibinfo{person}{Paul
  Denny}, \bibinfo{person}{Raymond Pettit}, \bibinfo{person}{Durell Bouchard},
  \bibinfo{person}{Dennis~J. Bouvier}, \bibinfo{person}{Brian Harrington},
  \bibinfo{person}{Amir Kamil}, \bibinfo{person}{Amey Karkare},
  \bibinfo{person}{Chris McDonald}, \bibinfo{person}{Peter-Michael Osera},
  \bibinfo{person}{Janice~L. Pearce}, {and} \bibinfo{person}{James Prather}.}
  \bibinfo{year}{2019}\natexlab{}.
\newblock \showarticletitle{Compiler Error Messages Considered Unhelpful: The
  Landscape of Text-Based Programming Error Message Research}. In
  \bibinfo{booktitle}{\emph{Proceedings of the Working Group Reports on
  Innovation and Technology in Computer Science Education}} (Aberdeen,
  Scotland, {UK}) \emph{(\bibinfo{series}{ITiCSE-WGR '19})}.
  \bibinfo{publisher}{ACM}, \bibinfo{address}{NY, NY, USA},
  \bibinfo{pages}{177–210}.
\newblock
\showISBNx{9781450375672}
\urldef\tempurl%
\url{https://doi.org/10.1145/3344429.3372508}
\showURL{%
\tempurl}


\bibitem[Campbell et~al\mbox{.}(2014)]%
        {campbell2014syntax}
\bibfield{author}{\bibinfo{person}{Hazel~Victoria Campbell},
  \bibinfo{person}{Abram Hindle}, {and} \bibinfo{person}{Jos\'{e}~Nelson
  Amaral}.} \bibinfo{year}{2014}\natexlab{}.
\newblock \showarticletitle{Syntax Errors Just Aren't Natural: Improving Error
  Reporting with Language Models}. In \bibinfo{booktitle}{\emph{Proceedings of
  the 11th Working Conference on Mining Software Repositories}} (Hyderabad,
  India) \emph{(\bibinfo{series}{MSR 2014})}. \bibinfo{publisher}{Association
  for Computing Machinery}, \bibinfo{address}{New York, NY, USA},
  \bibinfo{pages}{252–261}.
\newblock
\showISBNx{9781450328630}
\urldef\tempurl%
\url{https://doi.org/10.1145/2597073.2597102}
\showDOI{\tempurl}


\bibitem[Denny et~al\mbox{.}(2021)]%
        {denny2021designing}
\bibfield{author}{\bibinfo{person}{Paul Denny}, \bibinfo{person}{James
  Prather}, \bibinfo{person}{Brett~A. Becker}, \bibinfo{person}{Catherine
  Mooney}, \bibinfo{person}{John Homer}, \bibinfo{person}{Zachary~C Albrecht},
  {and} \bibinfo{person}{Garrett~B. Powell}.} \bibinfo{year}{2021}\natexlab{}.
\newblock \showarticletitle{On Designing Programming Error Messages for
  Novices: Readability and Its Constituent Factors}. In
  \bibinfo{booktitle}{\emph{Proceedings of the 2021 CHI Conference on Human
  Factors in Computing Systems}} (Yokohama, Japan) \emph{(\bibinfo{series}{CHI
  '21})}. \bibinfo{publisher}{ACM}, \bibinfo{address}{NY, NY, USA}, Article
  \bibinfo{articleno}{55}, \bibinfo{numpages}{15}~pages.
\newblock
\showISBNx{9781450380966}
\urldef\tempurl%
\url{https://doi.org/10.1145/3411764.3445696}
\showURL{%
\tempurl}


\bibitem[Dy and Rodrigo(2010)]%
        {dy2010detector}
\bibfield{author}{\bibinfo{person}{Thomas Dy} {and}
  \bibinfo{person}{Ma.~Mercedes Rodrigo}.} \bibinfo{year}{2010}\natexlab{}.
\newblock \showarticletitle{A Detector for Non-Literal {Java} Errors}. In
  \bibinfo{booktitle}{\emph{Proceedings of the 10th Koli Calling International
  Conference on Computing Education Research}} (Koli, Finland)
  \emph{(\bibinfo{series}{Koli Calling '10})}. \bibinfo{publisher}{ACM},
  \bibinfo{address}{NY, NY, USA}, \bibinfo{pages}{118–122}.
\newblock
\showISBNx{9781450305204}
\urldef\tempurl%
\url{https://doi.org/10.1145/1930464.1930485}
\showDOI{\tempurl}


\bibitem[Finnie-Ansley et~al\mbox{.}(2022)]%
        {finnieansley2022robots}
\bibfield{author}{\bibinfo{person}{James Finnie-Ansley}, \bibinfo{person}{Paul
  Denny}, \bibinfo{person}{Brett~A. Becker}, \bibinfo{person}{Andrew
  Luxton-Reilly}, {and} \bibinfo{person}{James Prather}.}
  \bibinfo{year}{2022}\natexlab{}.
\newblock \showarticletitle{The Robots Are Coming: Exploring the Implications
  of {OpenAI} {Codex} on Introductory Programming}. In
  \bibinfo{booktitle}{\emph{Proceedings of the 24th Australasian Computing
  Education Conference}} (Virtual Event, Australia) \emph{(\bibinfo{series}{ACE
  '22})}. \bibinfo{publisher}{ACM}, \bibinfo{address}{New York, NY, USA},
  \bibinfo{pages}{10--19}.
\newblock
\showISBNx{9781450396431}
\urldef\tempurl%
\url{https://doi.org/10.1145/3511861.3511863}
\showDOI{\tempurl}


\bibitem[Finnie-Ansley et~al\mbox{.}(2023)]%
        {finnieansley2023myai}
\bibfield{author}{\bibinfo{person}{James Finnie-Ansley}, \bibinfo{person}{Paul
  Denny}, \bibinfo{person}{Andrew Luxton-Reilly},
  \bibinfo{person}{Eddie~Antonio Santos}, \bibinfo{person}{James Prather},
  {and} \bibinfo{person}{Brett~A. Becker}.} \bibinfo{year}{2023}\natexlab{}.
\newblock \showarticletitle{My {AI} Wants to Know if this will be on the Exam:
  Testing OpenAI's Codex on {CS2} Programming Exercises}. In
  \bibinfo{booktitle}{\emph{Australasian Computing Education Conference}}
  (Melbourne, VIC, Australia) \emph{(\bibinfo{series}{ACE '23})}.
  \bibinfo{publisher}{ACM}, \bibinfo{address}{NY, NY, USA},
  \bibinfo{numpages}{8}~pages.
\newblock
\urldef\tempurl%
\url{https://doi.org/10.1145/3576123.3576134}
\showDOI{\tempurl}


\bibitem[Gupta et~al\mbox{.}(2017)]%
        {gupta2017deepfix}
\bibfield{author}{\bibinfo{person}{Rahul Gupta}, \bibinfo{person}{Soham Pal},
  \bibinfo{person}{Aditya Kanade}, {and} \bibinfo{person}{Shirish Shevade}.}
  \bibinfo{year}{2017}\natexlab{}.
\newblock \showarticletitle{{DeepFix}: Fixing Common {C} Language Errors by
  Deep Learning}. In \bibinfo{booktitle}{\emph{Proceedings of the {AAAI}
  Conference on Artificial Intelligence}}, Vol.~\bibinfo{volume}{31}.
  \bibinfo{publisher}{Association for the Advancement of Artificial
  Intelligence}, \bibinfo{address}{San Francisco, CA, USA},
  \bibinfo{pages}{1345--1351}.
\newblock
\urldef\tempurl%
\url{https://doi.org/10.1609/aaai.v31i1.10742}
\showDOI{\tempurl}


\bibitem[Hindle et~al\mbox{.}(2012)]%
        {hindle2012naturalness}
\bibfield{author}{\bibinfo{person}{Abram Hindle}, \bibinfo{person}{Earl~T.
  Barr}, \bibinfo{person}{Zhendong Su}, \bibinfo{person}{Mark Gabel}, {and}
  \bibinfo{person}{Premkumar Devanbu}.} \bibinfo{year}{2012}\natexlab{}.
\newblock \showarticletitle{On the Naturalness of Software}. In
  \bibinfo{booktitle}{\emph{34th International Conference on Software
  Engineering ({ICSE})}}. \bibinfo{publisher}{{IEEE}},
  \bibinfo{address}{Zurich, CH}, \bibinfo{pages}{837--847}.
\newblock
\urldef\tempurl%
\url{https://doi.org/10.1109/ICSE.2012.6227135}
\showDOI{\tempurl}
\newblock
\shownote{{ISSN}: 1558-1225}.


\bibitem[Howard and Ruder(2018)]%
        {howard2018universallanguagemodelfinetuning}
\bibfield{author}{\bibinfo{person}{Jeremy Howard} {and}
  \bibinfo{person}{Sebastian Ruder}.} \bibinfo{year}{2018}\natexlab{}.
\newblock \bibinfo{title}{Universal Language Model Fine-tuning for Text
  Classification}.
\newblock
\newblock
\showeprint[arxiv]{1801.06146}~[cs.CL]
\urldef\tempurl%
\url{https://arxiv.org/abs/1801.06146}
\showURL{%
\tempurl}


\bibitem[Kohn(2019)]%
        {kohn2019error}
\bibfield{author}{\bibinfo{person}{Tobias Kohn}.}
  \bibinfo{year}{2019}\natexlab{}.
\newblock \showarticletitle{The Error Behind the Message: Finding the Cause of
  Error Messages in {Python}}. In \bibinfo{booktitle}{\emph{Proceedings of the
  50th {ACM} Technical Symposium on Computer Science Education}}.
  \bibinfo{publisher}{ACM}, \bibinfo{address}{NY, {NY}, {USA}},
  \bibinfo{pages}{524--530}.
\newblock


\bibitem[Kohn and Manaris(2020)]%
        {kohn2020tell}
\bibfield{author}{\bibinfo{person}{Tobias Kohn} {and} \bibinfo{person}{Bill
  Manaris}.} \bibinfo{year}{2020}\natexlab{}.
\newblock \showarticletitle{Tell Me What's Wrong: A {Python} {IDE} with Error
  Messages}.
\newblock In \bibinfo{booktitle}{\emph{Proceedings of the 51st {ACM} Technical
  Symposium on Computer Science Education}}. \bibinfo{publisher}{ACM},
  \bibinfo{address}{NY, {NY}, {USA}}, \bibinfo{pages}{1054--1060}.
\newblock
\showISBNx{978-1-4503-6793-6}
\urldef\tempurl%
\url{http://doi.org/10.1145/3328778.3366920}
\showURL{%
\tempurl}


\bibitem[Landis and Koch(1977)]%
        {landis1977measurement}
\bibfield{author}{\bibinfo{person}{J.~Richard Landis} {and}
  \bibinfo{person}{Gary~G. Koch}.} \bibinfo{year}{1977}\natexlab{}.
\newblock \showarticletitle{The Measurement of Observer Agreement for
  Categorical Data}.
\newblock \bibinfo{journal}{\emph{Biometrics}} \bibinfo{volume}{33},
  \bibinfo{number}{1} (\bibinfo{date}{Mar} \bibinfo{year}{1977}),
  \bibinfo{pages}{159--174}.
\newblock


\bibitem[Lau and Guo(2023)]%
        {lau2023ban}
\bibfield{author}{\bibinfo{person}{Sam Lau} {and} \bibinfo{person}{Philip
  Guo}.} \bibinfo{year}{2023}\natexlab{}.
\newblock \showarticletitle{From ``Ban It Till We Understand It'' to
  ``Resistance is Futile'': How University Programming Instructors Plan to
  Adapt as More Students Use AI Code Generation and Explanation Tools such as
  ChatGPT and GitHub Copilot}. In \bibinfo{booktitle}{\emph{Proceedings of the
  2023 ACM Conference on International Computing Education Research - Volume
  1}} (Chicago, {IL}, {USA}) \emph{(\bibinfo{series}{ICER '23})}.
  \bibinfo{publisher}{Association for Computing Machinery},
  \bibinfo{address}{New York, NY, USA}, \bibinfo{pages}{106–121}.
\newblock
\showISBNx{9781450399760}
\urldef\tempurl%
\url{https://doi.org/10.1145/3568813.3600138}
\showDOI{\tempurl}


\bibitem[Leinonen et~al\mbox{.}(2023)]%
        {leinonen2023using}
\bibfield{author}{\bibinfo{person}{Juho Leinonen}, \bibinfo{person}{Arto
  Hellas}, \bibinfo{person}{Sami Sarsa}, \bibinfo{person}{Brent Reeves},
  \bibinfo{person}{Paul Denny}, \bibinfo{person}{James Prather}, {and}
  \bibinfo{person}{Brett~A. Becker}.} \bibinfo{year}{2023}\natexlab{}.
\newblock \showarticletitle{Using Large Language Models to Enhance Programming
  Error Messages}. In \bibinfo{booktitle}{\emph{Proceedings of the 54th ACM
  Technical Symposium on Computer Science Education V. 1}} (Toronto ON, Canada)
  \emph{(\bibinfo{series}{SIGCSE 2023})}. \bibinfo{publisher}{ACM},
  \bibinfo{address}{NY, NY, USA}, \bibinfo{pages}{563–569}.
\newblock
\showISBNx{9781450394314}
\urldef\tempurl%
\url{https://doi.org/10.1145/3545945.3569770}
\showDOI{\tempurl}


\bibitem[Mahajan et~al\mbox{.}(2020)]%
        {mahajan2020recommending}
\bibfield{author}{\bibinfo{person}{Sonal Mahajan}, \bibinfo{person}{Negarsadat
  Abolhassani}, {and} \bibinfo{person}{Mukul~R. Prasad}.}
  \bibinfo{year}{2020}\natexlab{}.
\newblock \showarticletitle{Recommending {Stack} {Overflow} Posts for Fixing
  Runtime Exceptions Using Failure Scenario Matching}. In
  \bibinfo{booktitle}{\emph{Proceedings of the 28th ACM Joint Meeting on
  European Software Engineering Conference and Symposium on the Foundations of
  Software Engineering}} (Virtual Event, USA) \emph{(\bibinfo{series}{ESEC/FSE
  2020})}. \bibinfo{publisher}{Association for Computing Machinery},
  \bibinfo{address}{New York, NY, USA}, \bibinfo{pages}{1052–1064}.
\newblock
\showISBNx{9781450370431}
\urldef\tempurl%
\url{https://doi.org/10.1145/3368089.3409764}
\showDOI{\tempurl}


\bibitem[Marceau et~al\mbox{.}(2011)]%
        {marceau2011mind}
\bibfield{author}{\bibinfo{person}{Guillaume Marceau}, \bibinfo{person}{Kathi
  Fisler}, {and} \bibinfo{person}{Shriram Krishnamurthi}.}
  \bibinfo{year}{2011}\natexlab{}.
\newblock \showarticletitle{Mind Your Language: On Novices' Interactions with
  Error Messages}. In \bibinfo{booktitle}{\emph{Proceedings of the 10th SIGPLAN
  Symposium on New Ideas, New Paradigms, and Reflections on Programming and
  Software}} (Portland, Oregon, USA) \emph{(\bibinfo{series}{Onward! 2011})}.
  \bibinfo{publisher}{ACM}, \bibinfo{address}{NY, NY, USA},
  \bibinfo{pages}{3–18}.
\newblock
\showISBNx{9781450309417}
\urldef\tempurl%
\url{https://doi.org/10.1145/2048237.2048241}
\showDOI{\tempurl}


\bibitem[McCall and K{\"o}lling(2014)]%
        {mccall2014meaningful}
\bibfield{author}{\bibinfo{person}{Davin McCall} {and} \bibinfo{person}{Michael
  K{\"o}lling}.} \bibinfo{year}{2014}\natexlab{}.
\newblock \showarticletitle{Meaningful Categorisation of Novice Programmer
  Errors}. In \bibinfo{booktitle}{\emph{2014 {IEEE} {Frontiers} in Education
  Conference ({FIE}) Proceedings}}. \bibinfo{publisher}{IEEE},
  \bibinfo{address}{Madrid, Spain}, \bibinfo{pages}{1--8}.
\newblock


\bibitem[McCall and K{\"o}lling(2019)]%
        {mccall2019new}
\bibfield{author}{\bibinfo{person}{Davin McCall} {and} \bibinfo{person}{Michael
  K{\"o}lling}.} \bibinfo{year}{2019}\natexlab{}.
\newblock \showarticletitle{A New Look at Novice Programmer Errors}.
\newblock \bibinfo{journal}{\emph{ACM Transactions on Computing Education}}
  \bibinfo{volume}{19}, \bibinfo{number}{4} (\bibinfo{date}{July}
  \bibinfo{year}{2019}), \bibinfo{pages}{38:1--38:30}.
\newblock
\urldef\tempurl%
\url{https://doi.org/10.1145/3335814}
\showDOI{\tempurl}
\newblock
\shownote{\url{https://doi.org/10.1145/3335814}}.


\bibitem[Mozannar et~al\mbox{.}(2024)]%
        {mozannar2024realhumaneval}
\bibfield{author}{\bibinfo{person}{Hussein Mozannar}, \bibinfo{person}{Valerie
  Chen}, \bibinfo{person}{Mohammed Alsobay}, \bibinfo{person}{Subhro Das},
  \bibinfo{person}{Sebastian Zhao}, \bibinfo{person}{Dennis Wei},
  \bibinfo{person}{Manish Nagireddy}, \bibinfo{person}{Prasanna Sattigeri},
  \bibinfo{person}{Ameet Talwalkar}, {and} \bibinfo{person}{David Sontag}.}
  \bibinfo{year}{2024}\natexlab{}.
\newblock \bibinfo{title}{The RealHumanEval: Evaluating Large Language Models'
  Abilities to Support Programmers}.
\newblock
\newblock
\showeprint[arxiv]{2404.02806}~[cs.SE]


\bibitem[Nass and Moon(2000)]%
        {nass2000machines}
\bibfield{author}{\bibinfo{person}{Clifford Nass} {and}
  \bibinfo{person}{Youngme Moon}.} \bibinfo{year}{2000}\natexlab{}.
\newblock \showarticletitle{Machines and Mindlessness: Social Responses to
  Computers}.
\newblock \bibinfo{journal}{\emph{Journal of Social Issues}}
  \bibinfo{volume}{56}, \bibinfo{number}{1} (\bibinfo{year}{2000}),
  \bibinfo{pages}{81--103}.
\newblock
\showISSN{1540-4560}
\urldef\tempurl%
\url{https://doi.org/10.1111/0022-4537.00153}
\showDOI{\tempurl}


\bibitem[Nguyen et~al\mbox{.}(2024)]%
        {nguyen2024beginning}
\bibfield{author}{\bibinfo{person}{Sydney Nguyen},
  \bibinfo{person}{Hannah~McLean Babe}, \bibinfo{person}{Yangtian Zi},
  \bibinfo{person}{Arjun Guha}, \bibinfo{person}{Carolyn~Jane Anderson}, {and}
  \bibinfo{person}{Molly~Q Feldman}.} \bibinfo{year}{2024}\natexlab{}.
\newblock \bibinfo{title}{How Beginning Programmers and Code LLMs (Mis)read
  Each Other}.
\newblock
\newblock
\showeprint[arxiv]{2401.15232}~[cs.HC]


\bibitem[{OpenAI}(2024a)]%
        {openai2024finetuning}
\bibfield{author}{\bibinfo{person}{{OpenAI}}.}
  \bibinfo{year}{2024}\natexlab{a}.
\newblock \bibinfo{booktitle}{\emph{Fine-tuning - OpenAI API}}.
\newblock {OpenAI}.
\newblock
\urldef\tempurl%
\url{https://platform.openai.com/docs/guides/fine-tuning/example-count-recommendations}
\showURL{%
Retrieved 2024-07-21 from \tempurl}


\bibitem[{OpenAI}(2024b)]%
        {openai2024prompt}
\bibfield{author}{\bibinfo{person}{{OpenAI}}.}
  \bibinfo{year}{2024}\natexlab{b}.
\newblock \bibinfo{booktitle}{\emph{Prompt engineering - {OpenAI} {API}}}.
\newblock {OpenAI}.
\newblock
\urldef\tempurl%
\url{https://platform.openai.com/docs/guides/prompt-engineering/tactic-provide-examples}
\showURL{%
Retrieved 2024-07-21 from \tempurl}


\bibitem[Phung et~al\mbox{.}(2023)]%
        {phung2023generating}
\bibfield{author}{\bibinfo{person}{Tung Phung}, \bibinfo{person}{José
  Cambronero}, \bibinfo{person}{Sumit Gulwani}, \bibinfo{person}{Tobias Kohn},
  \bibinfo{person}{Rupak Majumdar}, \bibinfo{person}{Adish Singla}, {and}
  \bibinfo{person}{Gustavo Soares}.} \bibinfo{year}{2023}\natexlab{}.
\newblock \bibinfo{title}{Generating High-Precision Feedback for Programming
  Syntax Errors using Large Language Models}.
\newblock
\newblock
\showeprint[arxiv]{2302.04662}~[cs.PL]


\bibitem[Prather et~al\mbox{.}(2024)]%
        {prather2024widening}
\bibfield{author}{\bibinfo{person}{James Prather}, \bibinfo{person}{Brent
  Reeves}, \bibinfo{person}{Juho Leinonen}, \bibinfo{person}{Stephen MacNeil},
  \bibinfo{person}{Arisoa~S. Randrianasolo}, \bibinfo{person}{Brett Becker},
  \bibinfo{person}{Bailey Kimmel}, \bibinfo{person}{Jared Wright}, {and}
  \bibinfo{person}{Ben Briggs}.} \bibinfo{year}{2024}\natexlab{}.
\newblock \bibinfo{title}{The Widening Gap: The Benefits and Harms of
  Generative AI for Novice Programmers}.
\newblock
\newblock
\showeprint[arxiv]{2405.17739}~[cs.AI]
\urldef\tempurl%
\url{https://arxiv.org/abs/2405.17739}
\showURL{%
\tempurl}


\bibitem[Santos and Becker(2024)]%
        {santos2024not}
\bibfield{author}{\bibinfo{person}{Eddie~Antonio Santos} {and}
  \bibinfo{person}{Brett~A. Becker}.} \bibinfo{year}{2024}\natexlab{}.
\newblock \showarticletitle{Not the Silver Bullet: {LLM-enhanced} Programming
  Error Messages are Ineffective in Practice}. In
  \bibinfo{booktitle}{\emph{Proceedings of the 2024 Conference on United
  Kingdom \& Ireland Computing Education Research}} (Manchester, United
  Kingdom) \emph{(\bibinfo{series}{UKICER '24})}. \bibinfo{publisher}{ACM},
  \bibinfo{address}{New York, NY, USA}, Article \bibinfo{articleno}{5},
  \bibinfo{numpages}{7}~pages.
\newblock
\showISBNx{9798400711770}
\urldef\tempurl%
\url{https://doi.org/10.1145/3689535.3689554}
\showDOI{\tempurl}


\bibitem[Santos et~al\mbox{.}(2018)]%
        {santos2018syntax}
\bibfield{author}{\bibinfo{person}{Eddie~Antonio Santos},
  \bibinfo{person}{Hazel~Victoria Campbell}, \bibinfo{person}{Dhvani Patel},
  \bibinfo{person}{Abram Hindle}, {and} \bibinfo{person}{Jos{\'e}~Nelson
  Amaral}.} \bibinfo{year}{2018}\natexlab{}.
\newblock \showarticletitle{Syntax and {Sensibility}: Using Language Models to
  Detect and Correct Syntax Errors}. In \bibinfo{booktitle}{\emph{2018 {{IEEE}}
  25th {{International Conference}} on {{Software Analysis}}, {{Evolution}} and
  {{Reengineering}} ({{SANER}})}}. \bibinfo{publisher}{{IEEE}},
  \bibinfo{address}{Campobasso, Italy}, \bibinfo{pages}{311--322}.
\newblock


\bibitem[Santos et~al\mbox{.}(2023)]%
        {santos2023always}
\bibfield{author}{\bibinfo{person}{Eddie~Antonio Santos},
  \bibinfo{person}{Prajish Prasad}, {and} \bibinfo{person}{Brett~A. Becker}.}
  \bibinfo{year}{2023}\natexlab{}.
\newblock \showarticletitle{Always Provide Context: The Effects of Code Context
  on Programming Error Message Enhancement}. In
  \bibinfo{booktitle}{\emph{Proceedings of the ACM Conference on Global
  Computing Education Vol 1}} (Hyderabad, India) \emph{(\bibinfo{series}{CompEd
  2023})}. \bibinfo{publisher}{ACM}, \bibinfo{address}{New York, NY, USA},
  \bibinfo{pages}{147–153}.
\newblock
\showISBNx{9798400700484}
\urldef\tempurl%
\url{https://doi.org/10.1145/3576882.3617909}
\showDOI{\tempurl}


\bibitem[Shneiderman(1982)]%
        {shneiderman1982system}
\bibfield{author}{\bibinfo{person}{Ben Shneiderman}.}
  \bibinfo{year}{1982}\natexlab{}.
\newblock \showarticletitle{System Message Design: Guidelines and Experimental
  Results}.
\newblock In \bibinfo{booktitle}{\emph{Directions in Human/Computer
  Interaction}}, \bibfield{editor}{\bibinfo{person}{Albert Badre} {and}
  \bibinfo{person}{Ben Shneiderman}} (Eds.). \bibinfo{publisher}{Ablex
  Publishing Company}, \bibinfo{address}{Norwood, NJ}, Chapter~3,
  \bibinfo{pages}{55--77}.
\newblock


\bibitem[Vaithilingam et~al\mbox{.}(2022)]%
        {vaithilingam2022expectation}
\bibfield{author}{\bibinfo{person}{Priyan Vaithilingam},
  \bibinfo{person}{Tianyi Zhang}, {and} \bibinfo{person}{Elena~L. Glassman}.}
  \bibinfo{year}{2022}\natexlab{}.
\newblock \showarticletitle{Expectation vs. Experience: Evaluating the
  Usability of Code Generation Tools Powered by Large Language Models}. In
  \bibinfo{booktitle}{\emph{{CHI} Conference on Human Factors in Computing
  Systems Extended Abstracts}}. \bibinfo{publisher}{ACM}, \bibinfo{address}{NY
  NY, USA}, \bibinfo{pages}{1--7}.
\newblock


\bibitem[Widjojo and Treude(2023)]%
        {widjojo2023addressing}
\bibfield{author}{\bibinfo{person}{Patricia Widjojo} {and}
  \bibinfo{person}{Christoph Treude}.} \bibinfo{year}{2023}\natexlab{}.
\newblock \bibinfo{title}{Addressing Compiler Errors: {Stack Overflow} or Large
  Language Models?}
\newblock
\newblock
\showeprint[arxiv]{2307.10793}~[cs.SE]


\end{thebibliography}

\end{document}